\documentclass{emulateapj}

\usepackage{times}

\shorttitle{Large-scale shock surrounding 3C\,444}
\shortauthors{Croston et al.}

\begin{document}


\title{A large-scale shock surrounding a powerful radio galaxy}


\author{J. H. Croston\altaffilmark{1}, M. J. Hardcastle\altaffilmark{2}, B. Mingo\altaffilmark{2}, D. A. Evans\altaffilmark{3,4}, D. Dicken\altaffilmark{5}, R. Morganti\altaffilmark{6,7}, C. N. Tadhunter\altaffilmark{8}}
\altaffiltext{1}{School of Physics and Astronomy, University of Southampton,
  Southampton, SO17 1BJ, UK.; J.Croston@soton.ac.uk}
\altaffiltext{2}{School of Physics, Astronomy and Mathematics, University of
  Hertfordshire, Hatfield, Hertfordshire, AL10 9AB, UK}
\altaffiltext{3}{Harvard-Smithsonian Center for Astrophysics, 60 Garden Street, Cambridge, MA 02138, USA}
\altaffiltext{4}{Elon University, Elon, NC 27244, USA}
\altaffiltext{5}{Department of Physics and Astronomy, Rochester Institute of Technology, 84 Lomb Memorial Drive, Rochester, NY 14623, USA}
\altaffiltext{6}{Netherlands Institute for Radio Astronomy, Postbus 2, 7990 AA, Dwingeloo}
\altaffiltext{7}{Kapteyn Astronomical Institute, University of Groningen, P.O. Box 800, 9700 AV Groningen, The Netherlands}
\altaffiltext{8}{Department of Physics and Astronomy, University of Sheffield, Sheffield S3 7RH}



\begin{abstract}
We report the {\it Chandra} detection of a large-scale shock, on
scales of 200 kpc, in the cluster surrounding the powerful radio
galaxy 3C\,444 (PKS 2211$-$17).  Our 20-ks {\it Chandra} observation
allows us to identify a clear surface brightness drop around the outer
edge of the radio galaxy, which is likely to correspond to a
spheroidal shock propagating into the intracluster medium. We measure
a temperature jump across the surface brightness drop of a factor
$\sim 1.7$, which corresponds to a Mach number of $\sim 1.7$. This is
likely to be an underestimate due to the need to average over a fairly
large region when measuring the temperature of the post-shock gas. We
also detect clear cavities corresponding to the positions of the radio
lobes, which is only the second such detection associated with an FRII
radio galaxy. We estimate that the total energy transferred to the
environment is at least $8.2 \times 10^{60}$ ergs, corresponding to a
jet power of $>2.2 \times 10^{45}$ ergs s$^{-1}$ (assuming a timescale
based on the measured shock speed). We also compare the external
pressure acting on the lobes with the internal pressure under various
assumptions, and conclude that a significant contribution from protons
is required.
\end{abstract}





\section{Introduction}

The presence of large-scale shocks surrounding powerful radio galaxies
has long been a prediction from models of their evolution (e.g. Scheuer
1974; Kaiser \& Alexander 1997), and it was realised prior to the
launch of the {\it Chandra} X-ray observatory that such shocks in
radio-galaxy environments should be detectable with our current
generation of X-ray instruments (e.g. Heinz, Reynolds \& Begelman
1998). Deep {\it Chandra} observations have now firmly identified weak
shocks in the intracluster medium surrounding nearby low-power [FRI:
Fanaroff \& Riley (1974) class 1] sources, such as Perseus, Hydra A,
and M87 (e.g. Fabian et al.  2003; Nulsen et al. 2005a; Forman et
al. 2007), and a few examples of strong shocks have been found
surrounding smaller sources (e.g. Kraft et al. 2003; Croston et
al. 2007; Croston et al. 2009).

Recent work investigating radio galaxy impact with large samples
(e.g. Dunn et al. 2008; B\^{i}rzan et al. 2008; Cavagnolo et al. 2010)
has also tended to focus on lower power objects, for two major
reasons. Firstly their role as a means of feedback in galaxy evolution
is currently perceived to be more important, due to their higher
number density and ubiquity at the centers of ``cool-core'' clusters
(e.g. Cavagnolo et al. 2008). In addition, most X-ray cavity and weak
shock detections reported in the literature are associated with FRI
radio galaxies, due mainly to the relative scarcity of FRIIs in
the nearby clusters where cavities and shocks can easily be detected,
but also perhaps partly due to the presence of X-ray inverse-Compton
emission which dominates over thermal emission from the ICM for many
nearby FRII radio galaxies in poorer environments (e.g. Croston et
al. 2005; Kataoka \& Stawarz 2005).

Shocks surrounding powerful FRII radio galaxies have so far proved
somewhat elusive, although there is evidence for a large-scale shock
surrounding Cygnus A (Smith et al. 2002), as well as a weak shock
surrounding the intermediate FRI/II radio galaxy Hercules A (Nulsen et
al. 2005b). The investigation of shocks around FRII radio galaxies is
important, however, as they could play a role in galaxy feedback at
high redshift (e.g. Rawlings \& Jarvis 2004). Detections of
shocks and cavities associated with FRII radio galaxies offer an
independent means of investigating the conclusions of inverse-Compton
and environmental studies, which suggest that in most cases FRII radio
galaxies are close to equipartition between radiating particles and
magnetic field, and do not require a significant proton population for
pressure balance (e.g. Croston et al.  2005, Kataoka \& Stawarz 2005),
in constrast to the FRI population (e.g. Croston et al. 2008a). A few
FRIIs in rich cluster environments appear to deviate from this
conclusion (e.g. Belsole et al. 2007; Hardcastle \& Croston 2010),
which might indicate a relationship between particle content and
environment. It is important to establish whether such a relationship
exists in order to account correctly for the energetics of different
radio-galaxy populations within galaxy feedback models.

Here we report the detection of a large-scale shock surrounding the
weak-lined FR II radio galaxy 3C\,444 ($z = 0.153$, $L_{1.4 GHz} = 3.3
\times 10^{26}$ W Hz$^{-1}$), which is found at the center of the
cluster Abell 3847 ($L_{X} \sim 10^{44}$ erg s$^{-1}$ -- see below),
and present an investigation of its dynamics and
energetics. Throughout the paper we use a cosmology in which $H_0 =
70$ km s$^{-1}$ Mpc$^{-1}$, $\Omega_{\rm m} = 0.3$ and $\Omega_\Lambda
= 0.7$, corresponding to an angular scale of 2.66 kpc arcsec$^{-1}$ at
this redshift. Galactic absorption of $N_{\rm H} = 2.51 \times
10^{20}$ cm$^{-2}$ is assumed for all X-ray spectral fits. Spectral
indices $\alpha$ are defined in the sense $S_{\nu} \propto
\nu^{-\alpha}$. Reported errors are 1$\sigma$ for one interesting
parameter, except where otherwise noted.
\section{Observations and data analysis}
\label{dataan}
We observed 3C\,444 (PKS\,2211$-$17) with the {\it Chandra} ACIS-S
detector for 20 ks on 2009 March 20th, as part of a programme to
complete observations of the southern 2Jy sample of radio galaxies
(Tadhunter et al. 1993; Morganti et al. 1993). The observation was
taken in VFAINT mode to minimize the background level. The data were
reprocessed from the level 1 events file with {\sc ciao} 4.2 and CALDB
4.3, including VFAINT cleaning. The latest gain files were applied and
the 0.5-pixel randomization removed using standard techniques detailed
in the {\sc ciao} on-line
documentation\footnote{http://asc.harvard.edu/ciao/}. An inspection of
the lightcurve for our observation using the {\it analyze\_ltcrv}
script showed that there were no periods of high background level, and
so no additional GTI filtering was applied.  The radio analysis in
this paper is based on archival VLA observations at 1.4-GHz and 5 GHz
(program ID AD276). The data were calibrated and imaged within {\sc
aips} in the standard manner.

\begin{figure}
\begin{center}
\plotone{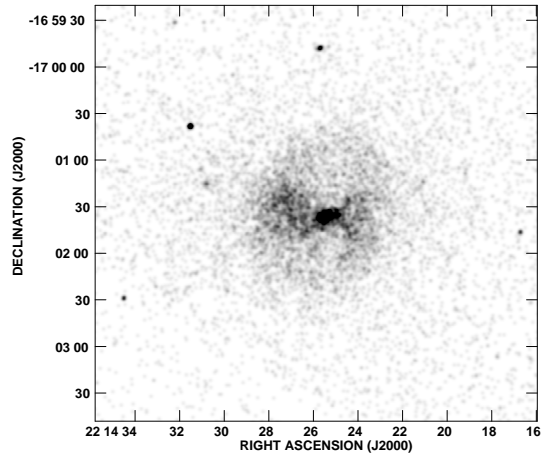}
\caption{A 0.5 - 5.0 keV image of the {\it Chandra} data, lightly
  smoothed with a Gaussian kernel of $\sigma = 3$ pix.}
\label{data}
\end{center}
\end{figure}

\begin{figure}
\begin{center}
\plotone{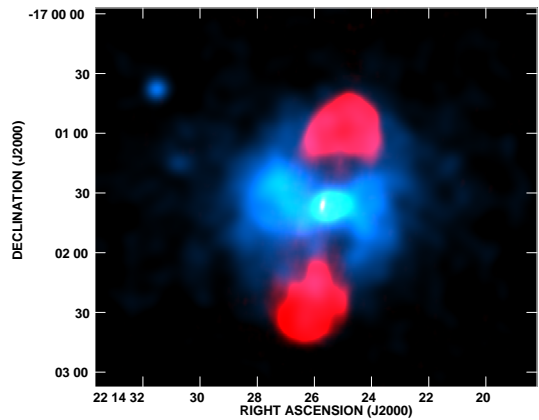}
\caption{An image made from the {\it Chandra} data (shown in blue,
  smoothed with a Gaussian of $\sigma= 15$ pix) and the 5-GHz VLA map (shown
  in red), indicating the relationship between the radio and X-ray
  structures, including cavities at the position of the radio lobes,
  and a sharp elliptical surface brightness drop surrounding the
  source.}
\label{overlay}
\end{center}
\end{figure}

We produced a 0.5 -- 5 keV filtered image from the {\it Chandra} data
to examine the X-ray emission associated with the radio galaxy and
environment, which is presented in Fig.~\ref{data}.  The relationship
between the X-ray and radio emission is shown in
Fig.~\ref{overlay}. Extended emission associated with the surrounding
galaxy cluster is clearly detected, with prominent cavities at the
positions of the two radio lobes. We note that this is only the second
detection of clear cavities associated with a radio galaxy that can be
unambiguously classified as an FRII based on its morphology (narrow
collimated jets and a compact hotspot in the northern lobe) and radio
luminosity.  The X-ray surface brightness also decreases noticeably
beyond an elliptical region enclosing the radio galaxy.

\begin{figure}
\plotone{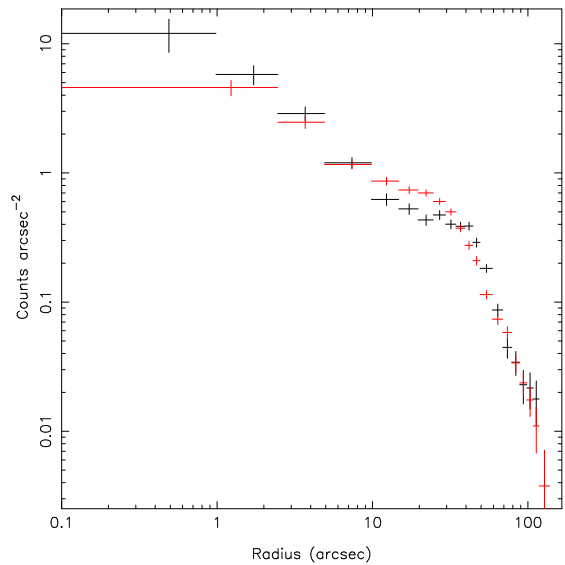}
\caption{X-ray surface brightness profiles of the emission surrounding
3C\,444. The emission was divided into 4 quadrants, with the North and
South quadrants combined to make a longitudinal profile (black) and the
East and West quandrants combined to make a transverse profile
(red). A sharp drop is present in the N-S direction at $\sim 45$
arcsec, while a less sharp drop is present at $\sim 30$ arcsec in the
E-W direction.}.
\label{sprof}
\end{figure}

We first extracted a global spectrum for the cluster (excluding
  the central 2$^{\prime\prime}$ and the radio lobes to avoid
  non-thermal contamination) and fitted an {\sc apec} model to
  characterize its global properties. We measure a global temperature
  of $kT = 3.5\pm0.2$ keV, and a bolometric X-ray luminosity of $1.0
  \times 10^{44}$ erg s$^{-1}$, consistent with a moderately rich
  cluster and in line with observed $L_{X}$--$T_{X}$ relations (e.g
  Pratt et al. 2009). The overall X-ray morphology is very regular,
  apart from the X-ray cavities, and as reported later
  (Table~\ref{spec} and Fig.~\ref{sprof}) we do observe a moderate
  decrease in central temperature and peak in surface brightness;
  however, the central density based on extrapolating from a beta
  model fit to the inner regions is slightly lower than the value of
  $h(z)^{-2}n_{e,0} > 4 \times 10^{-2}$ cm$^{-2}$ used to distinguish
  cool-core clusters by Pratt et al. (2009), hence it may not be
  a strong cool-core system.

\begin{figure}
\begin{center}
\plotone{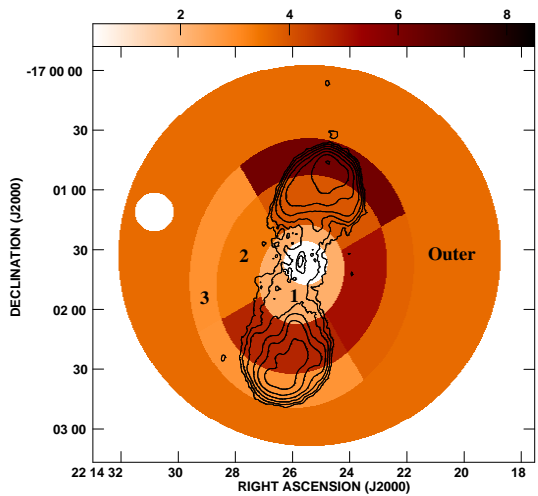}
\caption{Temperature map (in keV) produced using the results of the
  spectral fits in Table~\ref{spec} (regions as in Fig,~\ref{data}),
  with 5-GHz contours overlaid. White regions do not have measured
  temperatures. The small circle to the east is a contaminating point
  source excluded from the fits. The regions used in Table~\ref{spec} are indicated with annulus numbers/region names.}
\label{tmap}
\end{center}
\end{figure}

In order to investigate whether the observed surface brightness
decrease is associated with a shock, we extracted surface brightness
profiles and spectra in several regions of interest. In particular we
considered an elliptical region chosen to surround the bright X-ray
emission, which we divided into three elliptical annuli. Two of
  the annuli were further subdivided into quadrants to investigate
  temperature structure. Our spectral extraction regions are indicated
  in Fig.~\ref{tmap}. An elliptical annulus beyond the bright X-ray
region was also used (``Outer region'' in Table~\ref{spec}).
Fig.~\ref{sprof} shows the surface brightness profiles in the N-S and
E-W directions. In both directions the profile flattens in a way
  not typically seen in undisturbed cluster profiles (e.g. Cavagnolo
  et al. 2009), and then drops steeply. The steepening is sharpest in
  the N-S direction, where there is an abrupt turn-over at a distance
  of $\sim 45$ arcsec, consistent with the edge of the X-ray bright
  region shown in Fig.~\ref{overlay}. The location of the turn-over
  appears to be at the same distance from the nucleus in both the
  North and South direction.

\begin{deluxetable*}{llrrrrr}
\tablewidth{12cm} \tablecaption{Results of spectral modelling for
  cluster gas emission and radio lobes\tablenotemark{a}}
\tablehead{Region&Subregion/Model&Temperature
  (keV)&$\Gamma$&$\chi^{2}$&D.O.F.\\} \startdata Annulus
1&&$2.2\pm0.2$&&36.5&39\\ Annulus
2&N&$4.0^{+0.7}_{-0.6}$&&24.6&27\\ &E&$3.4^{+0.4}_{-0.3}$&&28.8&39\\ &S&$4.8^{+1.1}_{-0.8}$&&10.2&19\\ &W&$5.1^{1.1}_{-0.8}$&&31.4&28\\ Annulus
3&N&$6.2^{+2.2}_{-1.3}$&&11.3&19\\ &E&$2.9^{+0.6}_{-0.5}$&&8.3&14\\ &S&$2.9^{+1.1}_{-0.6}$&&1.2&6\\ &W&$3.8^{+1.6}_{-1.0}$&&4.5&7\\ Outer
region&&$3.6^{+0.5}_{-0.4}$&&58.8&62\\ Lobes&PL&&$1.78\pm0.04$&82.5&71\\ &{\sc apec}&$3.91\pm0.33$&&63.8&71\\ &PL
+ {\sc apec}\tablenotemark{b}&$3.7\pm0.5$&1.7\tablenotemark{c}&63.7&70\\ \enddata
\tablenotetext{a}{All spectral fits are in the energy range $0.4 - 7$
  keV, using an {\sc apec} model with abundance fixed at $0.3 \times$
  solar abundance.}  \tablenotetext{b}{The 0.4 - 7.0 keV unabsorbed
  flux in was measured to be $<3.1 \times 10^{13}$ ergs cm$^{-2}$
  s$^{-1}$ and $(5.2^{+0.7}_{-1.8}) \times 10^{-13}$ ergs cm$^{-2}$
  s$^{-1}$ for the PL and {\sc apec} components, respectively.}
\tablenotetext{c}{The power-law photon index tended to very low values
  of $\sim 0.3$, and so we fixed its value at a physically reasonable
  value for lobe inverse-Compton emission so as to test the possible
  presence of such a component.}  \tablenotetext{d}{The lower bound on
  the normalisation of the power-law component was consistent with
  zero.}
\label{spec}
\end{deluxetable*}

Spectra for the eight annular quadrants as well as the inner and outer
regions were fitted with the {\sc apec}\footnote{Astrophysical Plasmas
  Emission Code - http://cxc.harvard.edu/atomdb/} model. The results of this spectral fitting
are listed in Table~\ref{spec}. The abundance was poorly constrained
in free abundance fits, and so we report the results of fits with
abundance fixed at $0.3 \times$ solar abundance. A crude temperature
map using the same regions is shown in Fig.~\ref{tmap}. Spectral
results for the AGN nucleus will be presented in Mingo et al. (in
prep.). The nuclear count rate is insignificant compared to the
emission from the cluster gas, and so is not a contaminating source of
non-thermal emission for our analysis.

The results of spectral fitting to the radio lobe regions (extraction
regions chosen to encompass the extent of radio emission) are listed
in Table~\ref{spec}. It is clear from the detection of X-ray
cavities in our {\it Chandra} data that any lobe inverse-Compton
emission (cf. Croston et al. 2005) is weak relative to the cluster
X-ray emission; however, it is important to be certain that
non-thermal emission cannot be contaminating our spectral results. As
shown in Table~\ref{spec}, a thermal model was a significantly
better fit than a power-law model to the lobe regions; however, a
thermal + power-law model also gave a good fit with a significant
power-law contribution, corresponding to a 1-keV flux density of 9
nJy, which is a factor of $\sim 4$ higher than the IC prediction if
the source was at equipartition [based on modelling using the code of
Hardcastle et al. (1998)]. Such a departure from equipartition is
consistent with the results for the FRII population (e.g. Croston et
al. 2005); however, the presence of this spectral component is not
formally required by the data, therefore we cannot consider this an
inverse-Compton detection. Reassuringly, the temperature of the
thermal component in the model is not significantly affected by the
presence of a power-law contribution at this level. IC emission at a
higher level is unlikely: our previous work on large samples of FRII
radio galaxies has shown that the level of lobe IC is narrowly
distributed, with a mean level a factor 2--3 above the equipartition
prediction. We therefore conclude that our temperature measurements
are not affected by inverse-Compton contamination.

\section{Evidence for a shock surrounding 3C\,444}
As shown in Fig.~\ref{sprof}, the surface brightness drops by a factor
of $\sim 10$ between 45 and 70 arcsec from the AGN nucleus (or cluster
center). This corresponds roughly to a density jump of a factor of
$\sim 3$, consistent with the presence of a moderately strong
expanding shock wave surrounding the radio lobes of 3C\,444 at a
distance of $\sim 200$ kpc from the cluster centre.  Further evidence
for the presence of a shock comes from our spectral analysis: there is
a temperature jump of a factor $1.7\pm0.4$ at the edge of the North
lobe. Note that this temperature jump is likely to be an
underestimate: the relatively low count levels from our short
observation mean that large spectral extraction regions are required,
which are likely to average over cooler regions further from the shock
front. This temperature jump corresponds to a Mach number of
$1.7\pm0.4$, confirming that a shock is indeed present.  The Mach
number calculated is significantly lower than that implied by the
surface brightness jump, ${\cal M} \sim 3$. This may be because the
temperature jump is underestimated due to averaging the hottest
emission regions with cooler material further from the shock, or could
be partly due to the geometrical and uniform density assumptions. The
temperature structure around the South radio lobe is more ambiguous -
the temperature appears higher in a region somewhat closer to the
nucleus, and comparatively cool ($\sim 3$ keV) at the outer edge. The
reason for this is unclear, but could be due to significant asymmetry
in the temperature distribution of the environment and perhaps some
contribution from projection effects.

\section{Dynamics and energetic impact of 3C\,444}
To investigate further the source dynamics and energetic impact, we
estimated the external pressure surrounding the outer radio lobes, and
compared it with the internal radio lobe pressure under various
assumptions. We estimated the external pressure acting on the radio
lobes by considering the northern and southern quadrants of annuli 2
and 3, described above. We assumed each region had the geometry of an
ellipsoidal shell (subtracting the overlapping volume of the radio
lobe in each case) with a constant density contained within it and the
mean temperature from the two spectral regions corresponding to each
shell. Using the temperatures listed in Table~\ref{spec}, we
determined densities of $n_{\rm p} = 0.047$ and 0.036 cm$^{-3}$ for
the northern and southern quadrants, respectively, which correspond to
external pressures of $\sim 8.4 \times 10^{-12}$ Pa and $\sim 4.9
\times 10^{-12}$ Pa, respectively (we do not quote errors as the
uncertainty is dominated by systematic error due to the geometrical
assumptions and assumption of uniform density).  The internal lobe
pressures were determined under the assumption of no protons in three
different scenarios: (1) equipartition between magnetic field and
radiating particles, (2) a magnetic field strength $B = 0.7 B_{\rm
  eq}$, the median found for FRII radio galaxies in the
inverse-Compton study of Croston et al. (2005), and (3) the magnetic
field strength leading to inverse-Compton emission just below the
measured upper limit for the lobes of 9 nJy at 1 keV (see
Section~\ref{dataan}, above). These gave pressures for the North lobe
of $3.6 \times 10^{-13}$ Pa (scenario 1), $9 \times 10^{-13}$ Pa
(scenario 2), and $2.1 \times 10^{-12}$ Pa (scenario 3), and for the
South lobe of $2.3 \times 10^{-13}$ Pa (scenario 1), $6.7 \times
10^{-13}$ Pa (scenario 2), and $1.3 \times 10^{-12}$ Pa (scenario
3). A comparison with the external pressure measured above
($P_{ext}/P_{int} \sim 14 - 23$, $5-9$, and $4-6$, respectively, for
the three scenarios above) demonstrates that the lobes would be
underpressured in all of these scenarios, indicating that
non-radiating particles (e.g. protons) are likely to dominate the lobe
pressure.  An apparent pressure imbalance of this sort is commonly
found for low-power radio galaxies (e.g. Croston et al. 2008), but the
good agreement with equipartition magnetic fields from inverse-Compton
studies has previously been used to argue against a significant proton
population in FRII radio galaxies (e.g. Croston et
al. 2005). Environmental pressure comparisons for powerful FRII radio
galaxies (e.g. Croston et al. 2004; Belsole et al. 2007) indicate that
in general the requirements for proton content are smaller than in FRI
radio galaxies; however, FRII radio galaxies in the most rich
environments appear more proton-dominated (Belsole et al. 2007;
Hardcastle \& Croston 2010). This supports the argument of Croston et
al. (2008) that radio-galaxy particle content on large-scales is
related to radio morphology, which can indicate the radio jets'
ability to entrain material from the surroundings. The FRII radio
galaxies in rich environments, like 3C\,444, may therefore have a
higher proton content than those in poorer environments (e.g. Croston
et al. 2004).  

The thermal proton density in the heated region immediately
  surrounding the northern lobe edge was also used to estimate the
  total energetic impact of the radio galaxy. Assuming that the
  material in this region has been heated from the measured pre-shock
  temperature of 3.6 keV, the total ``excess'' energy in the hot
  region is $\sim 8.2 \times 10^{60}$ ergs. If the lobes are assumed
  slightly overpressured (e.g. $\sim 1.5 \times P_{ext}$), consistent
  with the observed shock, then the work done by the Northern lobe is
  $1.5 P_{ext} V \sim 1.4 \times 10^{60}$ ergs, a factor of $\sim 6$
  lower than the required heating. Note that as we have ignored any
heating in the E, W and S quadrants, the measured energy input could
therefore be considered a lower limit.  Finally, we used the measured
energy input from the radio galaxy to estimate the jet power. We
estimated the lobe inflation timescale by considering the sound
crossing time, as well as the inflation time if the expansion was
constant at the currently observed shock speed of ${\cal M} \sim
1.5$. The true evolution of the expansion speed cannot easily be
estimated: the lobe pressure would have been higher at earlier times,
but so would the external pressure closer to the cluster center. We
determine a sound-crossing timescale of $t_{cs} = 1.8 \times 10^{8}$
years, implying a jet power of $P_{jet} \sim 1.4 \times 10^{45}$ erg
s$^{-1}$. The ${\cal M}=1.5$ timescale is $1.2 \times 10^{8}$ years,
implying a jet power of $P_{jet} = 2.2 \times 10^{45}$ erg
s$^{-1}$. The sound crossing time is a conservative upper limit to the
source age, and our energy estimate is a lower limit due to
considering only the work done by the North lobe, and so we consider
the associated jet power a lower limit. The 1.4-GHz radio luminosity
of 3C\,444 is $L_{1.4GHz} = 3.3 \times 10^{26}$ W Hz$^{-1}$, and so
our estimated mechanical power is consistent, within the scatter, with
the relation of B\^{i}rzan et al. (2008). We note that as far as
  we are aware this is only the second jet power estimate from X-ray
  environmental observations, after Cygnus A, for a radio galaxy that
  can be classed unambigously as an FRII based on both radio
  morphology and luminosity. Our results are therefore important for
  constraining the poorly known high-luminosity end of the relation
  between jet power and radio luminosity.

\section{Conclusions}
We have identified a 200-kpc scale shock in the cluster environment of
the radio galaxy 3C\,444, based on a sharp surface brightness drop-off
and significant temperature increases at the edges of the radio
lobes. We infer that the radio galaxy has injected a total energy
$\sim 4$ times more than the work done by the radio lobes assuming
pressure balance, which implies that estimates of FRII radio galaxy
energy input based simply on $P$d$V$ work will be underestimates due
to the role of shock heating. We estimate a total energy input from
the expanding radio galaxy of $8.2 \times 10^{60}$ ergs, which implies
a jet power of $(1.4 -2.2) \times 10^{45}$ erg s$^{-1}$. Finally,
  based on a comparison of the external pressure with the internal
  pressure under various assumptions, we conclude that a significant
  proton contribution to the internal pressure is required, which is
  unexpected for an FRII galaxy given the good agreement of
  inverse-Compton measurements with the minimum energy condition for
  the population as a whole. However, we conclude that this is
  consistent with results for a handful of other powerful radio
  galaxies in rich environments, and suggests a strong role for
  environmental interactions in determining particle and energy
  content.

\acknowledgments

JHC acknowledges support from the South-East Physics Network (SEPnet).
MJH thanks the Royal Society for support via a University Research
Fellowship.

\end{document}